# Spin-glass behavior and pyroelectric anomalies in a new lithium-based oxide, Li$_3$FeRuO$_5$

Sanjay Kumar Upadhyay, P.L. Paulose, Kartik K Iyer, and E.V. Sampathkumaran
*Tata Institute of Fundamental Research, Homi Bhabha Road, Colaba, Mumbai 400005, India*

ABSTRACT

The results of dc and ac magnetization, heat-capacity, $^{57}$Fe Mössbauer spectroscopy, dielectric, pyroelectric current and isothermal magneto-capacitance measurements on a recently reported lithium-rich layered oxide, Li$_3$FeRuO$_5$, related to LiCoO$_2$-type (rhombohedral, space group $R\bar{3}m$), are presented. The results reveal that the compound undergoes spin-glass freezing at 15 K. There is a peak around 34 K in pyroelectric data, which can not be attributed to ferroelectricity, but to the phenomenon of thermally stimulated depolarization current. As revealed by magnetocapacitance data above and below magnetic ordering temperature, magnetic and electric dipoles appeared to be coupled, thereby offering evidence for magnetodielectric coupling.

**Introduction**

In an attempt to find new cathode materials for Li-ion batteries derived from layered oxides, $LiCoO_2$ and $Li_2MnO_3$, the synthesis of a rock salt related oxide, $Li_3FeRuO_5$, was reported by Laha et al [1] recently. This oxide was reported to order antiferromagnetically below 15 K. However, there is an intrinsic crystallographic disorder in this material (*see below*) and so one would expect that the exchange interaction must be sufficiently random that a long range magnetic order with a periodicity of magnetic structure would not be favoured. This is the primary motivation with which we have taken up this oxide for detailed magnetic investigations. The results establish that this oxide indeed exhibits spin-glass characteristics. In addition, we have carried out dielectric and pyro-electric studies, the results of which are also reported in this article.

The rhombohedral crystal structure (space group $R\bar{3}m$) in which the compound forms is related to that of cation-ordered rock-salt superstructure oxides, forming in $LiCoO_2$-type structure [2]. $LiCoO_2$ is made up of cubic close-packed oxygen array, with Li and Co ions occupying independent octahedral sites; the planes containing $LiO_2$ and $CoO_2$ alternate along [111] direction. In the compound under investigation, it was found that 3a site is mostly occupied by Li atom, while 3b site is occupied by Fe and Ru atom (almost equally). A fraction (<2%) of Fe and Ru goes to 3a site and around 20% of Li occupies 3b site. If we ignore a mixing of 2% of Fe and Ru at 3a site, the structure essentially consists of $LiO_6$ octahedra alternating with $(Li_{1/5}Fe_{2/5}Ru_{2/5}O_6)$ layer, running along the c-direction as shown in figure 1.

**Experimental details**

The polycrystalline $Li_3FeRuO_5$ was prepared by the conventional solid state route. The stoichiometric amounts of high purity (>99.9%) $Li_2CO_3$, $Fe_2O_3$ and $RuO_2$ were mixed together in an agate mortar, following by calcination at 925 $^0C$ for 12 hours. The heat-treated powder after grinding was then pre-sintered at 1025 $^0C$ for 12 hours. Final sintering was done at 1100 $^0C$ for 12 hours. X-ray diffraction (XRD) measurement was carried out with Cu $K_\alpha$ ($\lambda$ =1.54 $A^0$) and the pattern obtained is shown in figure 2. Within the detection limit (<2%) of the XRD technique, the prepared sample was found to be the single phase (rhombohedra, $R\bar{3}m$ space group). The experimental data was subjected to Rietveld refinement by using Fullprof programme [3]. The disorder among Li, Fe and Ru, as reported in Ref. 1, was taken into account, while refining. The lattice parameters are in good agreement with the literature [1].

Temperature ($T$) dependent dc magnetization ($M$) studies were carried out with the help of a commercial SQUID magnetometer (Quantum Design, USA) and ac magnetic susceptibility ($\chi$) studies with different frequencies ($\nu$= 1.3, 13, 133, and 1333 Hz) with a ac field of 1 Oe were also carried out with the same magnetometer. $^{57}Fe$ Mössbauer effect studies were carried out on the powder using a conventional constant acceleration spectrometer with a $^{57}Co$ source. The velocity calibration was carried out using a α-Fe absorber, and the typical linewidth obtained in our spectrometer was 0.28 mm/s. The isomer shift (IS) reported is relative to α-Fe. Heat-capacity (C) studies were carried with a commercial Physical Properties Measurements System (Quantum Design, USA). The same system was used to measure complex dielectric permittivity using an Agilent E4980A LCR meter with a home-made sample holder with several frequencies (1 kHz to 100 kHz) and with a bias voltage of 1 V; this sample holder was also used for pyroelectric studies with Keithley 6517B electrometer by poling at 100 K with different electric fields. Unlike otherwise stated, all the measurements were performed for zero-field-cooled condition (ZFC, from 330 K) of the specimen.

**Results**
**Dc magnetization**

Figure 3a shows inverse dc $\chi(T)$ of $Li_3FeRuO_5$ measured in the presence of ($H=$) 5 kOe for the ZFC protocol. Inverse $\chi$ curve was fitted by using the modified Curie-Weiss law in the temperature interval of 55- 175 K, accounting for 0.067 emu/mol due to magnetic impurity. The value of the effective moment obtained from this fitting is about 6.54 $\mu_B$ per formula unit which is very close to that expected (6.55 $\mu_B$) for trivalent Fe (S=5/2; high-spin) and tetra-valent Ru (S=1, low-spin) ions. If one has to derive the value of the magnetic moment from the higher temperature region, there is a need to suppress the impurity contribution by measuring at much higher fields. Therefore, we measured $\chi(T)$ in the presence of a high field of 30 kOe; the value of the effective moment from the Curie-Weiss fitting in the range 250-350 K turns out to be the same. The value of the paramagnetic Curie-temperature is found to be -18 K, indicating dominance of antiferromagnetic correlations. These findings are in good agreement with those reported by Laha et. al. [1] from the measurements with 50 Oe in the temperature range of 30 - 100 K. We have obtained the curves with a low-field of 100 Oe as well (figure 3a, inset). While ZFC curve qualitatively resembles that obtained with 5 kOe, the field-cooled (FC) curve deviates from this curve below 300 K, as observed in Ref. 1, attributable to traces of unreacted iron oxide. There is no downfall below 15 K in FC curve, but there is a tendency to flatten unlike in ZFC curve, which is a characteristic feature of spin-glasses.

We have measured isothermal magnetic hysteresis loops at selected temperatures (1.8 and 10 K) in the magnetically ordered state (see figure 3b). The curves show weak hysteretic behavior at low fields, but there is no evidence for saturation at high fields (even till 140 kOe, not shown here). Therefore, ferromagnetism is ruled out in this material. Viewed together with ZFC-FC $\chi(T)$ behavior, this *M(H)* behaviour can be attributed to spin-glass freezing. In order to ascertain spin-glass freezing, we performed other magnetic and heat-capacity measurements, the results of which are described below. Finally, we have also shown the *M(H)* curve for 300 K in the inset of figure 3b. The plot is linear, characteristic of paramagnetism, except for the fact that there is a weak upturn close to zero-field, attributable to magnetic impurity. The value of the moment extrapolated to zero field from the high-field linear region for this 300K-plot is less than 0.05$\mu_B$, and this value is consistent with only less than 2% of, say, $Fe_3O_4$.

**Heat capacity**

Figure 3c shows the plot of *C(T)* below 125 K obtained in the absence of external magnetic field as well as in 10 kOe. There is no evidence for any λ-anomaly (characterizing ferromagnetism or antiferromagnetism) down to 1.8 K in *C(T)*. This result is in favour of spin-glass freezing. A shoulder in the plot of *C/T* appears as a manifestation of spin-glass freezing (see figure 3c). The in-field curve overlaps with that obtained in zero field without any feature attributable to field-induced magnetic ordering. We could not obtain the values for electronic term to heat-capacity and the Debye temperature, as we find it difficult to see a wide linear range in the plot of *C/T* versus $T^2$ below 20 K due to the presence of the broad weak-shoulder in *C/T*. It is worth mentioning that the heat-capacity value at 300 K is about 238 J/mol K, which is in good agreement with the value of Dulong-Petit law for a molecule containing 10 atoms (249 J/mol K).

**Ac susceptibility**

Figure 4a shows the real ($\chi'$) and imaginary parts ($\chi''$) of ac $\chi$ behavior. It is obvious that, following the increase with a lowering of temperature, there is a peak in both these parts around 15 K agreeing with the dc magnetization behavior. $\chi'$ and $\chi''$ data reveal frequency dispersion which is a characteristic feature of spin glass freezing. We also measured the ac $\chi$ in different magnetic fields (10 and 20 kOe) and the frequency dependence of the peak is completely suppressed, that is, the curves overlap for different ν with a dramatic reduction in the values at the peaks. In fact, $\chi''$ is featureless in the high-field curves (and hence not shown in the figure). These findings are supportive of spin-glass ordering in zero field.

We have analysed the ac $\chi$ results with the power law, $\tau/\tau_0 = (T_f/T_g - 1)^{-zv}$ [4]. Here, $\tau$ represents the observation time (1/2πν), $\tau_0$ is the microscopic relaxation time, $T_g$ is the spin-glass transition temperature, $T_f$ corresponds to freezing temperature for a given observation time and zv is the critical exponent. For the feature around 15 K, we obtained $T_g \approx$ 14 K, zv ≈ 6.59 and $\tau_0 \approx$ 8.8x$10^{-12}$ sec. For a conventional spin glasses, the zv value falls in the range ~4-13, and $\tau_0$ value ranges between $10^{-10}$ and $10^{-13}$ s. It is clear that the values of $\tau_0$ obtained are in agreement with that for the conventional spin glasses [5].

**Isothermal remnant magnetization ($M_{IRM}$)**

We measured $M_{IRM}$ at four selected temperatures, 1.8, 10, 40 and 150 K. The specimen was zero-field-cooled to desired temperature, and then a field of 5 kOe was applied. After waiting for some time, the field was switched off, and then $M_{IRM}$ was measured as a function of time (*t*). We find that $M_{IRM}$ attained insignificantly small values within seconds of reducing the field to zero at 40 and 150 K; however, it decays slowly with *t* at other temperatures, as shown in figure 4b. This behavior of $M_{IRM}$ offers support to spin-glass dynamics. The curves were fitted to a stretched exponential form of the type $M_{IRM}(t) = M_{IRM}(0)[1+A\exp(-t/\tau)^{1-n}]$, where A and n are constants and $\tau$ here is the relaxation time. It is found that the value of n at 1.8 K and 10 K are 0.5 and 0.65 respectively. The values of relaxation times $\tau$ are ~2100 sec at 1.8 K and ~870 sec for 10 K. These values are in fact in agreement with that reported for spin-glasses [6].

**Memory effect in dc magnetization**

In order to look for 'memory' effect, characterizing spin-glasses [7, 8], we obtained $\chi(T)$ curves in different ways. In addition to ZFC curve in the presence of 100 Oe without a long wait at any temperature (which is a reference curve), we have obtained a ZFC curve after waiting at two temperatures 10 and 40 K for 3 hours each. We obtained the difference between these two curves and plotted the same as $\Delta M$ versus *T* in figure 4c. It is distinctly clear that there is a clear 'dip' at 10 K, and there is no dip at 40 K. This conclusively establishes spin-glass freezing in this compound.

**Waiting time dependence of dc magnetization**

We looked for aging effects [9, 10] in dc magnetization at 10 K, which is below $T_g$, the temperature characterizing the onset of spin-glass phase. For this purpose we have followed two different protocols: In ZFC protocol, we cooled the sample to the desired temperature, waited for a certain period of time, switched on a dc field of 100 Oe and measured the increase of *M* as a function of time. In the FC protocol, the specimen was cooled in 100 Oe and, after waiting for a certain period, the decay of *M* was measured as a function of time, after the field was switched off. The curves thus obtained for four waiting times ($t_w$=300, 1800, 3600 and 7200 s) are shown in figure 5. It is obvious from this figure that there is a systematic behaviour in FC as well as in ZFC aging, as a function of waiting time. In the ZFC plots, there is a gradual decrease in the magnetization value with waiting time. In FC aging, there is a gradual increase in the magnetization value as we increase the waiting from 300 sec to 7200 sec. Such a behaviour is an inherent property of frustrated spin systems. The aging behaviour is thus consistent with spin-glass dynamics.

**$^{57}$Fe Mössbauer effect measurements**

Mössbauer spectroscopy showed a paramagnetic quadrupolar split doublet at 298 K (Fig. 6a). However, the line width observed is 0.4 mm/s which is considerably larger than that of natural Fe in our set-up (0.28 mm/s). This could be attributed to the crystalline site disorder as was shown from the structural analysis. From a least square fit of the spectrum, we obtained isomer shift of 0.36 mm/s and quadrupolar splitting of 0.5 mm/s at 298 K. Both these values are consistent with Fe in 3+ state with S=5/2. The spectra show paramagnetic quadrupolar split doublet line down to 32 K and the quadrupolar splitting was unchanged. Well below 32 K, characteristic signature of magnetic hyperfine splitting is observed, but a resolved sextet could be obtained only at 16 K as can be readily seen from the figure. The absence of magnetic hyperfine splitting at higher temperatures also establishes that the fraction of other impurity oxides of Fe, say, $Fe_2O_3$ and $Fe_3O_4$, which are known to be in a long range magnetic ordered state, say, at 32 K, must be negligible. A good fit to the spectra were possible only with multiple hyperfine fields and the distribution of this field is shown in Fig. 6b. The 4.2 K spectrum could be fit using 6 terms in narrow region around 490 kOe. The average magnetic hyperfine field at 4.2 K is 490 kOe representative for $Fe^{3+}$ ions in high-spin configuration. The distribution is very broad above 4.2 K pointing to a strong inhomogeneous magnetic state. All these results establish that there is a significant crystallographic disorder.

**Complex permittivity and pyroelectric behaviour**

Complex permittivity and pyroelectric current ($I_{pyro}$) studies also reveal interesting behaviour well below and above $T_g$ (~15 K). Dielectric constants ($\varepsilon'$) and the loss factor (tan$\delta$) are shown in figure 7a below 100 K measured with a frequency of 100 kHz. Beyond 100 K, tan$\delta$ increases dramatically and therefore extrinsic contributions tend to dominate. The observation we would like to stress is that both $\varepsilon'$ and tan$\delta$ undergo a gradual increase with $T$ from 1.8 K, without any apparent peak or any other feature. Therefore, we rule out the presence of ferroelectricity below 150 K in this compound.

However, $I_{pyro}$ as a function of $T$ exhibits a distinct feature (measured with a poling voltage of 200 V corresponding to 4.16 kV/cm of the electric field). That is, the plot (figure 7b) shows a peak at about 34 K for a rate of warming of temperature ($dT/dt$) of 2 K/min. The peak gets inverted when poled by a reverse field as mentioned above, though the magnitude at the peak gets lowered. We have performed pyroelectric current measurements for different $dT/dt$. The results (see figure 7c) reveal that the peak in fact shifts to higher temperatures with increasing $dT/dt$, for instance, to ~36 K and ~37 K for 4 K/min and 6 K/min respectively. Such a rate dependence is not expected for ferroelectric transitions. This rate dependence can be explained in terms of 'thermally stimulated depolarization current (TSDC)' [11-13], that is, random trapping of mobile careers, which screen the applied electric field with a lowering of temperature, and persisting for a very long time after removal of the electric field. These carriers get released thermally, which appears as a peak in $I_{pyro}$ We have also obtained the behaviour of $I_{pyro}$ in the presence of a dc magnetic field of 50 kOe and we find (see Fig. 7b) that the intensity of the peak in the plot is slightly suppressed and shifted to a higher temperature by few degree Kelvin. Clearly magnetic field couples to these trapped electric dipoles.

In order to explore magneto-dielectric coupling (MDE), we have performed isothermal $\varepsilon'$ vs $H$ measurements at different temperature (2, 5, 10, 15, 20 and 40 K) up to 140 kOe. Figure 8 shows the MDE in the form $\Delta\varepsilon'$, where $\Delta\varepsilon' = [((\varepsilon'(H) - \varepsilon'(0))/\varepsilon'(0)]$, obtained with a measuring frequency of 100 kHz. From this figure, it is clear that there is a distinct difference in the magnitudes of $\Delta\varepsilon'$, say at 120 kOe, when the temperature is varied across $T_g$. That is, the value of $\Delta\varepsilon'$ gets relatively larger in the spin glass state, with respect to the values in the paramagnetic state, thereby establishing enhanced magnetodielectric coupling in the magnetically ordered state. One would also like mention here that, in the temperature regime of interest, the values of $tan\delta$ are in the range of 0.002-0.008, establishing highly insulating behaviour of the material and hence the observed MDE is an intrinsic effect.

**Conclusions**

The material, $Li_3FeRuO_5$, which was very recently synthesized for the first time [1] is investigated by thermal and magnetic studies as well as by $^{57}Fe$ Mössbauer spectroscopy. We find that, in this material, magnetic susceptibility obtained by ZFC protocol exhibits a peak near 15 K, while FC-$\chi$ curve tends to flatten at low temperatures. At the same temperature, there is a peak in ac $\chi$, which exhibits a frequency dependence in zero field, which is suppressed by applying a dc magnetic field, typical of canonical spin-glasses. Isothermal remnant magnetization exhibits slow decay with time, and 'aging' (waiting-time dependence) as well as 'memory' properties in dc magnetization as a function of temperature are also observed. λ-anomaly near 15 K in heat capacity expected for antiferromagnetic or ferromagnetic structure is also smeared. These findings conclusively establish that this compound undergoes spin-glass freezing below 15 K in this compound. The finding that Mössbauer spectra in the magnetically ordered state show multiple sextet patterns supports the conclusion from of the analysis of the XRD data that there is a significant crystallographic disorder, with different Fe ions experiencing different internal hyperfine fields, apparently due to local chemical environmental effects. We therefore conclude that intrinsic crystallographic disorder resulting in random exchange interaction is responsible for spin-glass freezing. Other findings we report here are: Magnetodielectric coupling gets relatively stronger as one enters spin-glass phase; complex dielectric permittivity and pyro-electric data rule out ferroelectricity; and there is a distinct evidence for thermally stimulated depolarization current phenomenon in this material around 34 K, which is presumably due to trapping of charges induced by crystallographic disorder.

Figure captions

Fig 1
Crystal structure of $Li_3FeRuO_5$ viewed along the [100] direction ignoring crystallographic disorder at 3a site as mentioned in the text.

Fig 2
X-ray diffraction pattern of $Li_3FeRuO_5$ along with the Reitveld refinement and derived parameters.

Fig 3
For $Li_3FeRuO_5$, (a) inverse magnetic susceptibility measured with 5 kOe, is plotted with a line fitted with the Curie-Weiss equation in the range 75-175 K, with the curves obtained with 100 Oe for both ZFC and FC conditions of measurements, in the inset, (b) isothermal magnetization hysteresis loops at 1.8 and 10 K with the *M(H)* curve for 300 K in the inset. And (c) Heat-capacity of $Li_3FeRuO_5$ in zero field as well as in 10 kOe.

Fig 4
For $Li_3FeRuO_5$, (a) real and imaginary parts of ac susceptibility measured with various frequencies (1.3, 13, 133 and 1333 Hz), with the arrows showing the direction in which the peaks shift with increasing frequency. The data points have been omitted for the sake of clarity and only the lines through data points are shown. (b) Isothermal remnant magnetization at 1.8, 10 and 40 K, and (c) the difference in magnetization curves, ΔM, obtained with and without waiting at 10 and 40 K. In figure (a), ac data in the presence of dc magnetic fields of 10 and 20 kOe are also shown.

Fig 5
Dc magnetization as a function of time (waiting time dependence or aging experiments) for (a) zero-field-cooled (ZFC) and (b) field-cooled (FC) protocols as described in the text for $Li_3FeRuO_5$ for 10 K. Arrows are drawn to identify to show how the curves shift with increasing waiting time.

Fig 6
For $Li_3FeRuO_5$, (a) $^{57}$Fe Mössbauer spectra and (b) Probability distribution as a function of hyperfine field, at various selected temperatures.

Fig 7
For $Li_3FeRuO_5$, temperature dependence of (a) dielectric constant (ε′) and (in the inset) loss factor (tanδ) measured with 100 kHz, (b) pyroelectric current, $I_{pyro}$, obtained by increasing the temperature at the rate of 2K/min after poling by 4.16 kV/cm and -4.16 kV/cm; and (c) $I_{pyro}$ as a function of *T* for different rates of change of $T$, after poling with 4.16 kV/cm, for $Li_3FeRuO_5$. In (b), the curve obtained in a field of 50 kOe is also included.

Fig 8
Isothermal magnetodielectric curves presented in the form of $\Delta\varepsilon'$ for selected temperatures, as described in text.

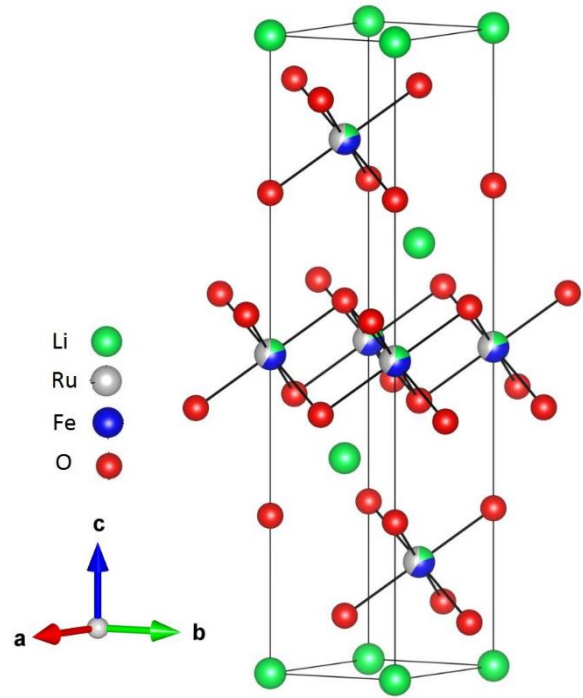

Fig 1

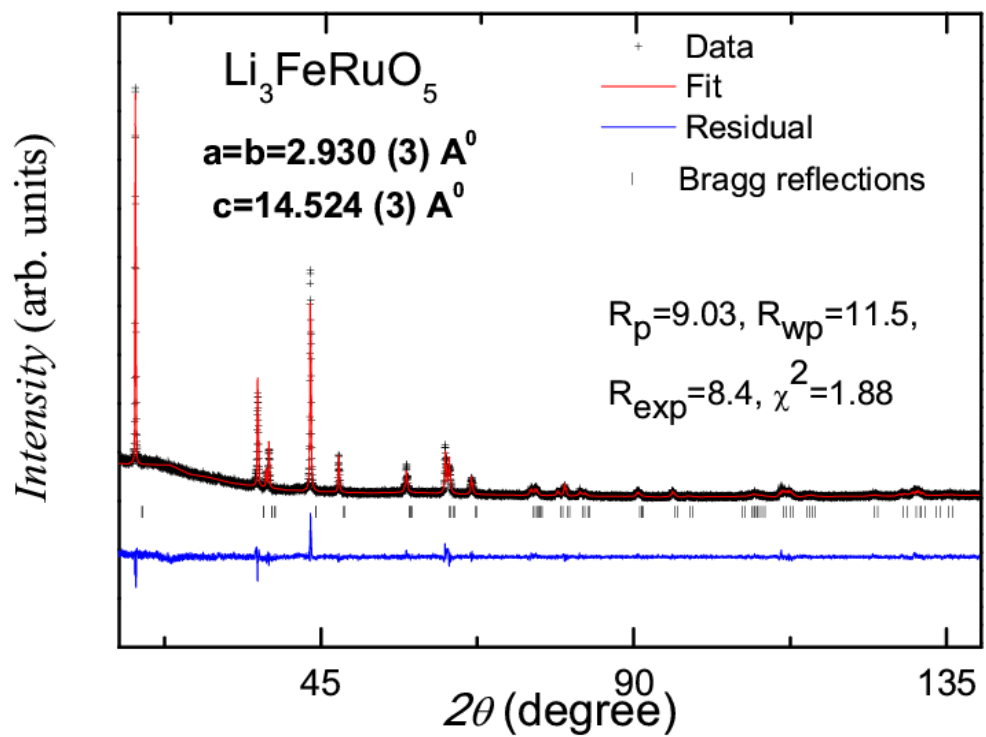

Fig 2

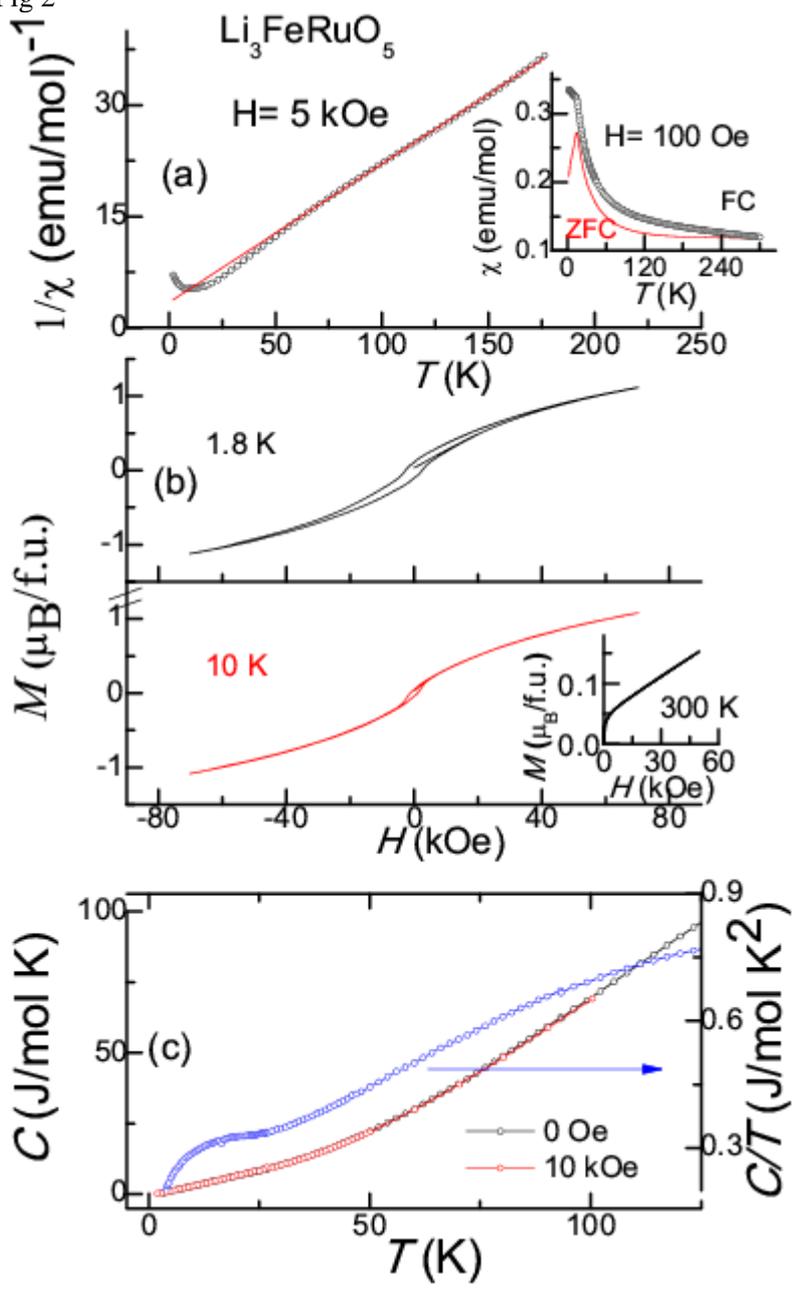

Fig 3

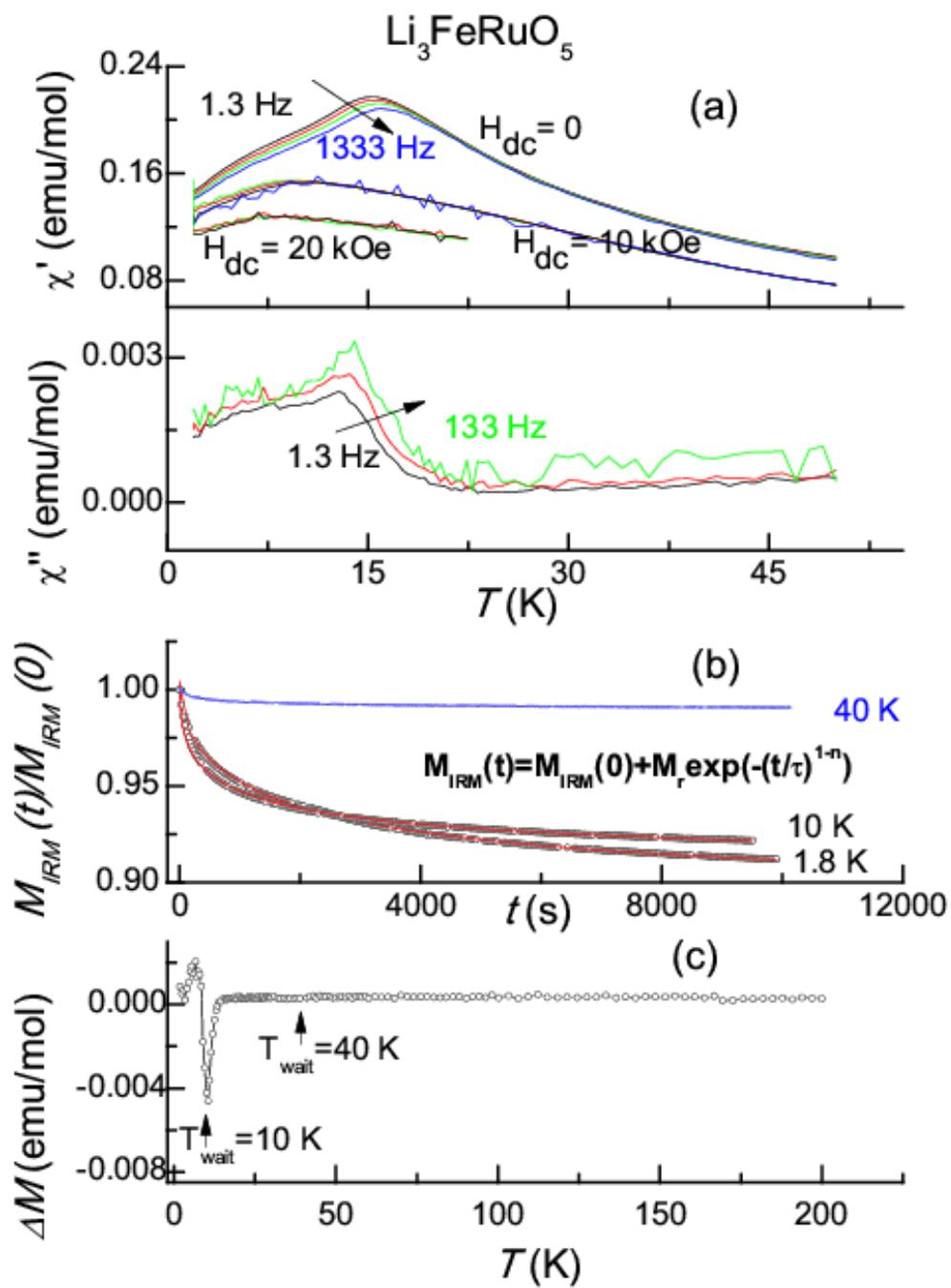

Fig 4

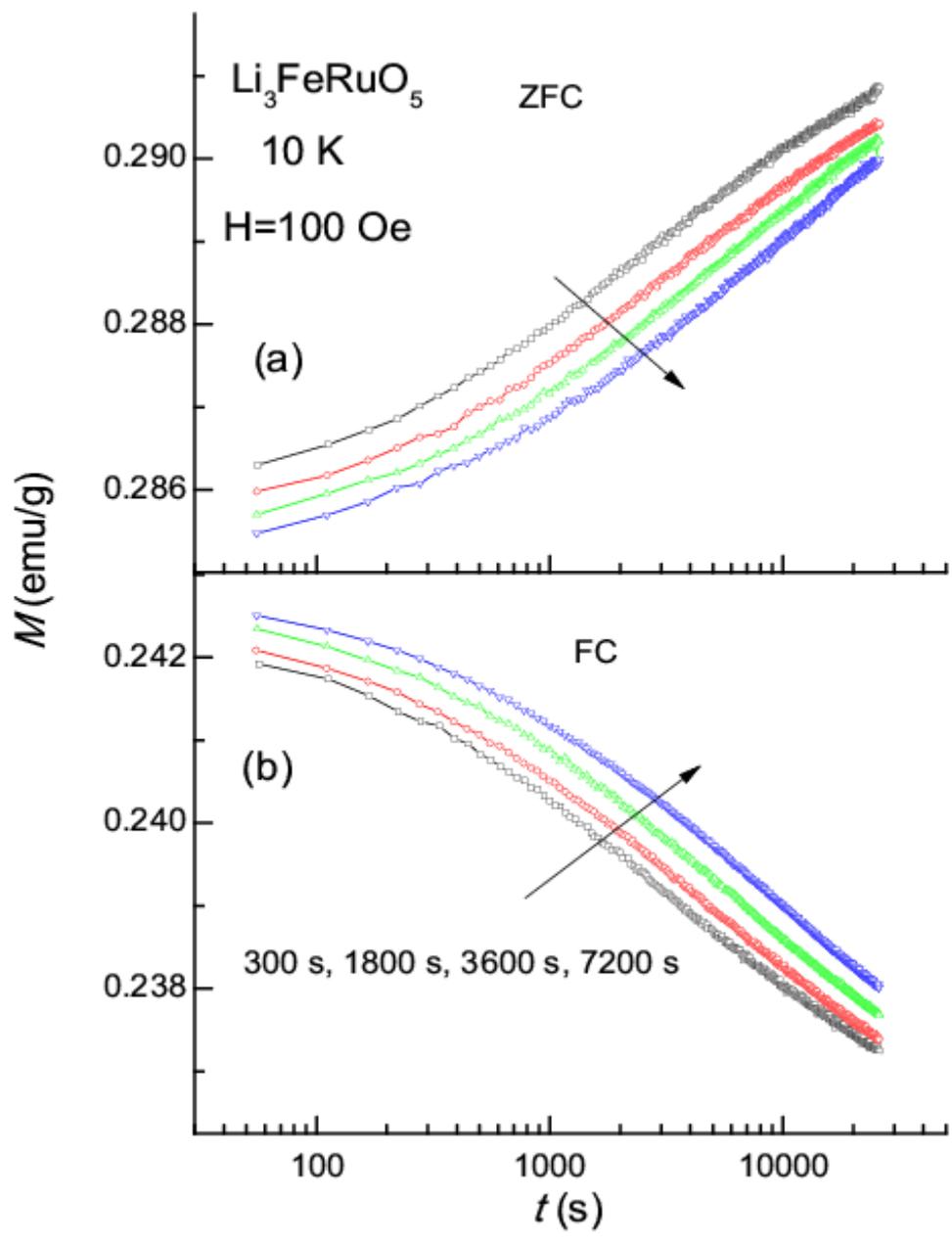

Fig 5

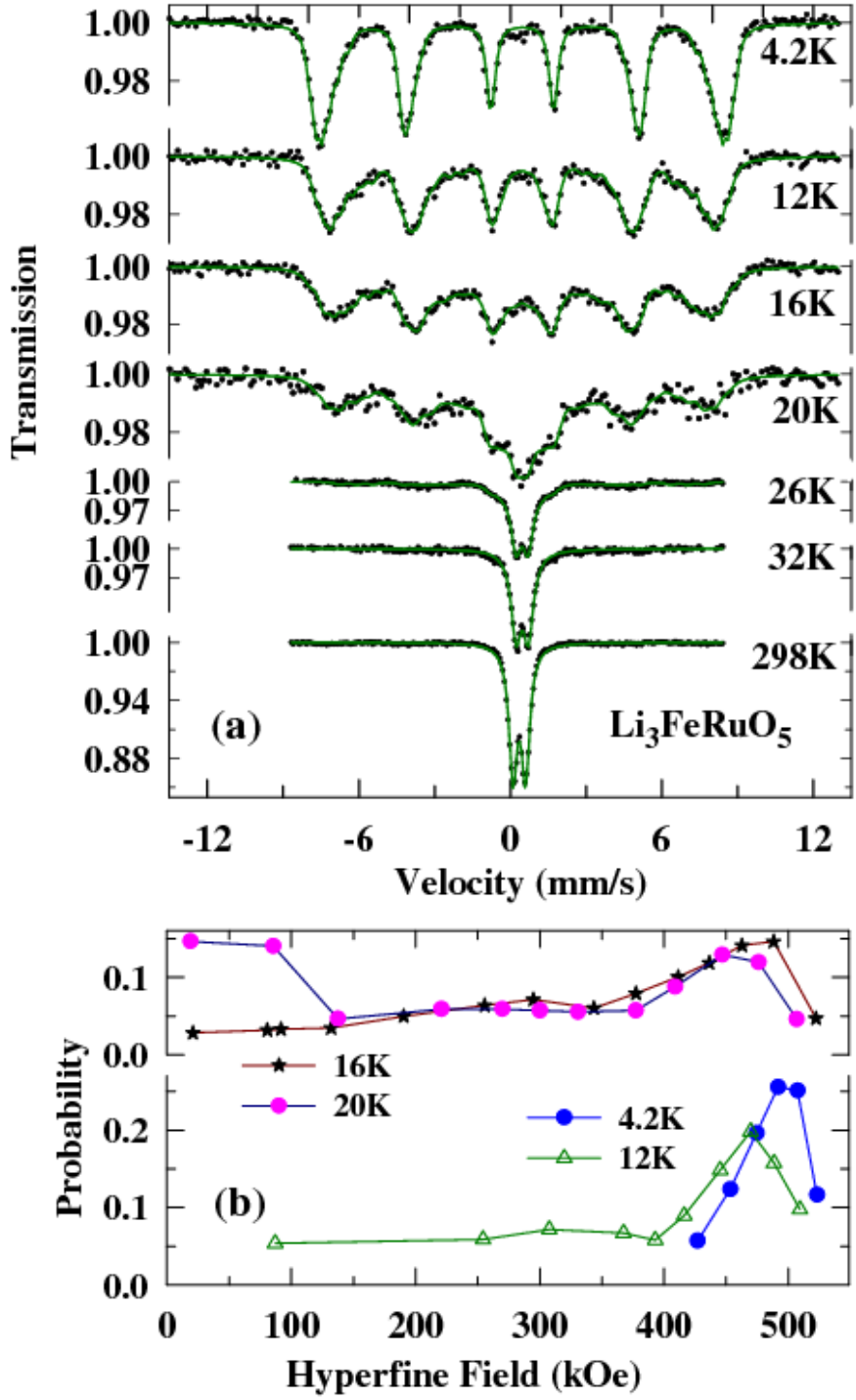

Fig 6

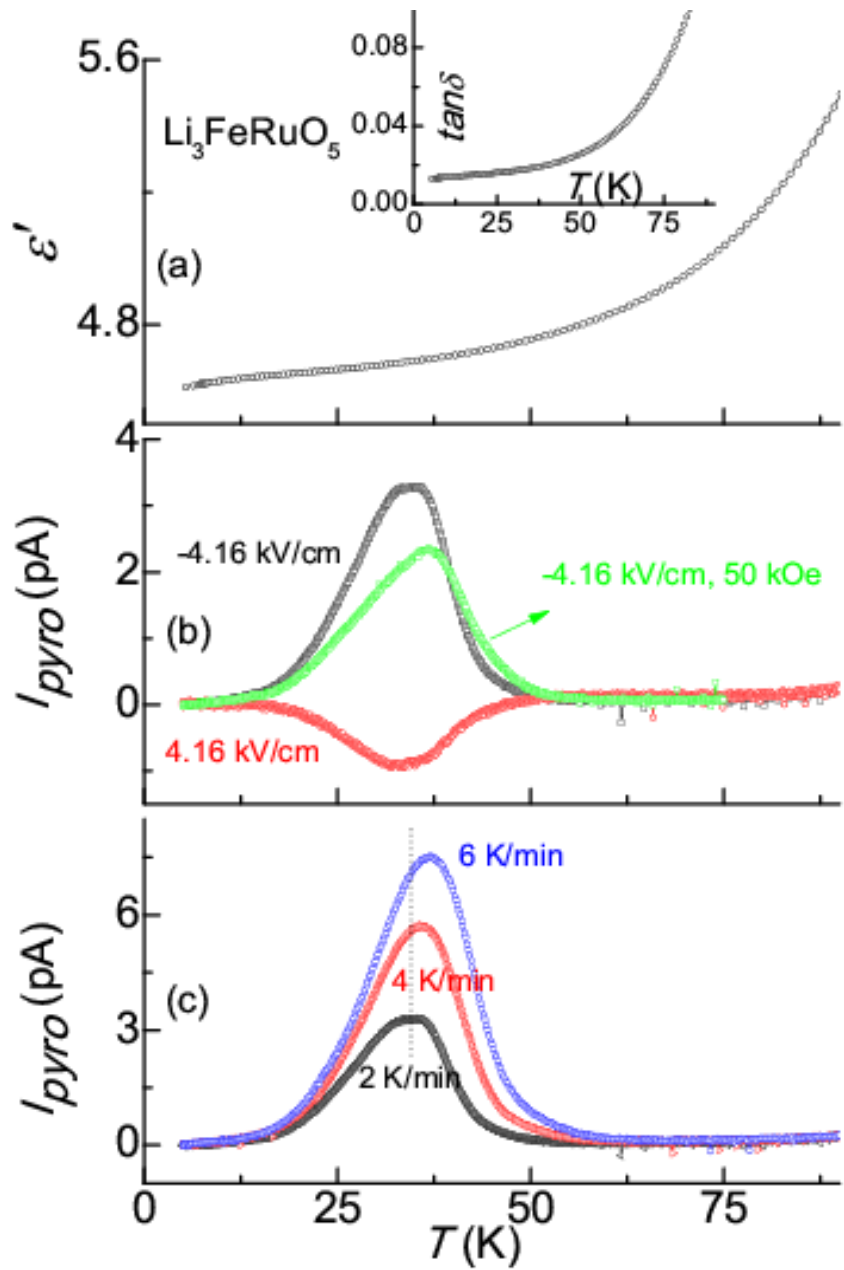

Fig 7

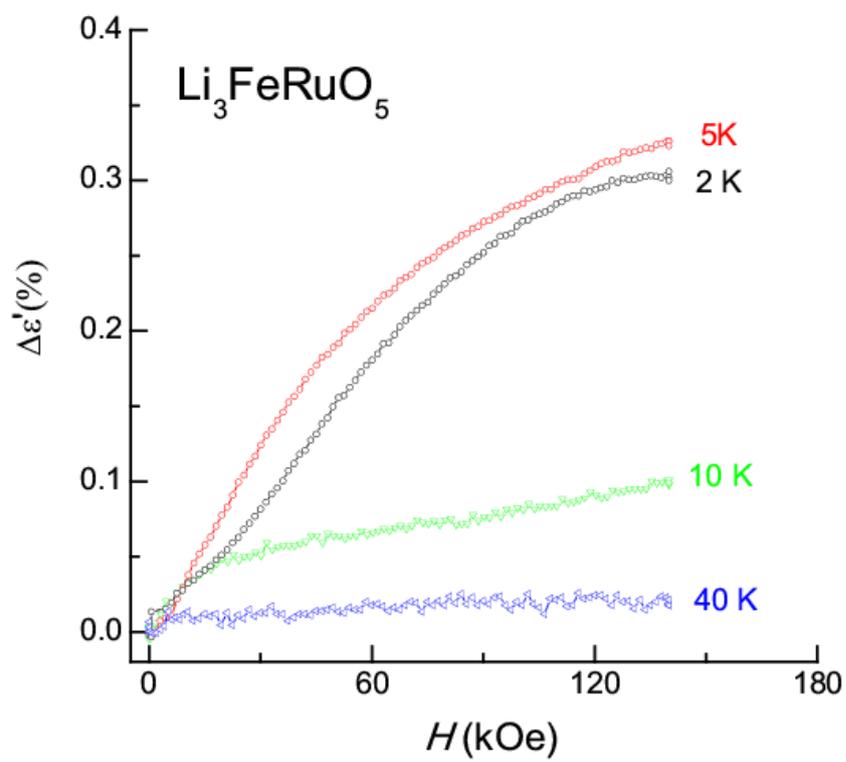

Fig 8